 \documentclass[authoryear,preprint,12pt]{elsarticle}
\usepackage{lineno}

\usepackage[table]{xcolor}

\usepackage{amssymb}
\usepackage{array}
\newcolumntype{C}[1]{>{\centering\let\newline\\\arraybackslash\hspace{0pt}}m{#1}}
\usepackage{longtable}
\usepackage{pdflscape}

 \usepackage{amsthm}
 \usepackage{amsmath}

\usepackage{booktabs}
\usepackage{tabularx}

\usepackage{multirow}
\usepackage{caption}
\usepackage{subcaption}
\usepackage{tabularx}
\usepackage{bm}

\usepackage{pgf, tikz} 
\usetikzlibrary{arrows,automata,fit}
\usetikzlibrary{shapes}

\usepackage{booktabs}
\usepackage{hyperref}

\journal{}

\begin{document}
\begin{frontmatter}



\title{Understanding support for AI regulation: A Bayesian network perspective}

\author[label1]{Andrea Cremaschi}
\author[label1]{Dae-Jin Lee}
 \author[label1]{Manuele Leonelli}
 \affiliation[label1]{organization={School of Science and Technology, IE University, Madrid}, country = {Spain}}


\begin{abstract}
As artificial intelligence (AI) becomes increasingly embedded in public and private life, understanding how citizens perceive its risks, benefits, and regulatory needs is essential. To inform ongoing regulatory efforts such as the European Union’s proposed AI Act, this study models public attitudes using Bayesian networks learned from the nationally representative 2023 German survey Current Questions on AI. The survey includes variables on AI interest, exposure, perceived threats and opportunities, awareness of EU regulation, and support for legal restrictions, along with key demographic and political indicators. We estimate probabilistic models that reveal how personal engagement and techno-optimism shape public perceptions, and how political orientation and age influence regulatory attitudes. Sobol indices and conditional inference identify belief patterns and scenario-specific responses across population profiles. We show that awareness of regulation is driven by information-seeking behavior, while support for legal requirements depends strongly on perceived policy adequacy and political alignment. Our approach offers a transparent, data-driven framework for identifying which public segments are most responsive to AI policy initiatives, providing insights to inform risk communication and governance strategies. We illustrate this through a focused analysis of support for AI regulation, quantifying the influence of political ideology, perceived risks, and regulatory awareness under different scenarios.
\end{abstract}




\end{frontmatter}

\section{Introduction}

Artificial intelligence (AI) is no longer a speculative technology: it is actively shaping how decisions are made, services are delivered, and risks are managed across nearly every domain of life. As AI systems become increasingly embedded in public infrastructure and private platforms, questions about how society should govern these technologies have gained new urgency. At the heart of this debate lies the public: how individuals perceive the risks and benefits of AI, how much they trust regulatory institutions, and whether they support legal constraints. These attitudes shape not only the democratic legitimacy of AI deployment, but also the political feasibility of governance frameworks \citep{li2020chb, chiarini2024}. Across Europe and beyond, policymakers are seeking to balance innovation with accountability, often amid incomplete or contested public consensus. The question of how people engage with AI has thus become central to both empirical research and policy design \citep{montag2023, dsit2024wave4}.

The European Union’s proposed AI Act provides a timely institutional backdrop. Structured around a tiered, risk-based approach to regulation \citep{kop2021, eprs2022}, the Act seeks to prohibit certain applications outright while tightly overseeing others deemed “high-risk.” Although widely praised for its ambition, the Act also raises questions about public understanding and perceived legitimacy \citep{laux2023}. Early evidence suggests that awareness of the regulation is uneven and that support varies with individuals’ broader beliefs about AI’s societal impact, their exposure to information, and political orientation \citep{magistro2025, gur2024}. Yet we still lack a coherent framework for understanding how these factors interact to shape public attitudes.

Existing research has identified many relevant predictors of AI-related beliefs, including age, education, ideology, and trust in institutions \citep{bergdahl2023, dong2024, yarovenko2024}. However, most studies use linear or additive models that estimate average effects, often overlooking the conditional relationships among variables and the underlying structure of belief systems \citep{cremaschi2025, schiavo2024}. As a result, we know relatively little about how people assemble their views: how, for example, exposure to information interacts with techno-optimism or how policy evaluations mediate support for regulation. Nor do most studies offer a transparent way to simulate how attitudes might shift under different scenarios or in different segments of the population.

This paper addresses gaps by modeling public attitudes toward AI using Bayesian networks (BNs)\citep{pearl1988probabilistic} learned from a nationally representative German survey conducted in 2023. Our approach captures the joint structure of beliefs across political, psychological, and informational dimensions, enabling us to explore how attitudes toward AI regulation are shaped by deeper patterns of perception and engagement. We find that support for legal requirements depends strongly on whether people view existing regulations as adequate, while awareness of those regulations is primarily driven by interest and information-seeking behavior. We also show that risk-oriented and opportunity-oriented respondents exhibit distinct logics of belief formation: one more ideologically structured, the other more technocratic. By making these relationships explicit, our model provides a data-driven foundation for designing more targeted communication strategies and responsive governance frameworks.

\section{Literature Review}

Public attitudes toward AI have been studied from diverse disciplinary perspectives, highlighting the roles of sociodemographics, ideology, trust, and regulatory awareness. While these factors are interdependent, existing research models them in isolation, using approaches that overlook the conditional structure of belief systems. The next sections review key predictors of AI-related attitudes and identify methodological gaps our study aims to address.

\subsection{Sociodemographic Drivers of Attitudes}

A broad literature investigates the factors shaping public perceptions of AI and automation. Demographic characteristics such as age, gender, and education are consistently associated with distinct attitudinal profiles: younger individuals and those with higher education levels tend to express greater optimism toward AI, while older and less educated respondents are more skeptical or fearful \citep{bergdahl2023, dsit2024wave4, li2020chb}. These associations have been observed across national contexts and survey instruments. Gender effects also emerge, with men typically expressing more confidence in AI and women showing greater concern about potential risks or inequalities.

However, demographics alone do not explain the complexity of AI perceptions. Psychological constructs such as technological self-efficacy, anxiety, and institutional trust have been shown to mediate the effects of background characteristics \citep{montag2023, schiavo2024}. In particular, AI-specific anxiety (encompassing concerns about bias, transparency, and control) has emerged as a distinct dimension that influences both expectations and policy preferences. Trust in government, science, and the media further modulates attitudes, with high-trust individuals more likely to accept the integration of AI in public life.

Political ideology also plays a central role. Research has shown that support for AI and automation is often polarized along left–right lines, especially when regulation or labor concerns are salient \citep{gur2024, yang2025}. Importantly, trust and ideology interact: political identity influences perceptions of institutional legitimacy, which in turn shapes how information about AI is received and evaluated. Cross-national comparisons reveal that these patterns are contingent on cultural and economic context. For instance, studies from Southern Europe and Latin America have found that fears about AI-driven job loss are particularly acute in low-trust environments, where technological change is perceived as externally imposed or unregulated \citep{dong2024, yarovenko2024, wilde2025}.

\subsection{Regulatory Preferences and Legal Frameworks}

Public support for AI regulation is generally high but conditional. Studies show that support tends to increase when individuals are made aware of specific risks, such as misuse in surveillance, discrimination in automated decisions, or the erosion of human agency \citep{dsit2024wave4, chiarini2024}. Regulatory support is also stronger when the perceived domain of application involves public goods, such as health care, education, or democratic participation. Yet this support is not uniform: ideological priors, institutional trust, and personal experience with technology all influence whether individuals see regulation as protective, burdensome, or symbolic.

Awareness of regulatory efforts, such as the EU AI Act, also varies across population segments. Recent studies suggest that public knowledge of such initiatives is limited, and that support depends not only on factual awareness but on how the regulation is framed and evaluated \citep{magistro2025}. Some individuals support legal restrictions in principle but view current institutional efforts as either inadequate or excessive; others may reject specific regulations while expressing trust in broader governance processes. These findings suggest that regulatory preferences are best understood as embedded within multidimensional belief systems, rather than as direct functions of exposure or ideology alone.

\subsection{Modeling Challenges and Interpretability}

Most empirical studies of public attitudes toward AI use linear regression, logistic models, or structural equation modeling to test directional hypotheses \citep{cremaschi2025, schiavo2024}. While effective for identifying average effects, these methods can obscure conditional dependencies, mediated paths, or interaction patterns that may be central to belief formation. Moreover, they typically rely on predefined structures, limiting the ability to discover emergent associations from data.

Recent work on interpretable machine learning emphasizes the value of transparent, flexible models in high-stakes domains \citep{rudin2019}. Probabilistic graphical models, including BNs, provide a structured way to capture the joint distribution of beliefs while retaining interpretability. These models allow for both exploratory learning and formal inference, enabling researchers to assess how beliefs propagate under hypothetical conditions or within subpopulations. As such, they are well suited to the analysis of complex public opinion systems\citep{cugnata2016bayesian} like those surrounding AI and its governance.

We contribute to this literature by applying BNs to the study of AI-related public attitudes using nationally representative European survey data. Our BN approach makes the conditional structure of belief systems explicit, modeling how perceptions, political orientation, and information exposure interact to shape awareness and support for AI regulation. This allows for transparent scenario-based inference and targeted policy insights, addressing both methodological and substantive gaps in existing research.

\section{Materials and Methods}

\begin{table*}
\footnotesize
\centering
\scalebox{0.62}{
\renewcommand{\arraystretch}{1.05}
\begin{tabular}{p{3.5cm}p{4cm}p{6.5cm}p{6cm}}
\toprule
\textbf{Group} & \textbf{Variable} & \textbf{Description (Question or Theme)} & \textbf{Levels} \\
\midrule
\multirow{5}{*}{\textbf{Demographics}} 
& Sex & Respondent's gender & Female, Male \\
& Age & Respondent's age group & 14--29, 30--44, 45--59, 60+ \\
& Education & Highest level of completed education & Low, Medium, High \\
& Income & Self-reported income category & Low, Medium, High, Not reported \\
& Municipality & Size of municipality of residence & $<$5k, 5k--19k, 20k--99k, 100k--499k, 500k+ \\
\midrule
\multirow{5}{*}{\textbf{Exposure}} 
& InterestAI & How interested are you in information about AI?  & Not at all, Less strongly, Strongly, Very strongly \\
& InformedAI & How well informed do you feel about AI?  & Very poor, Rather poor, Rather good, Very good \\
& MediaAI & Have you recently seen/heard anything about AI in the media?  & Yes, No \\
& FriendsAI & Have you discussed AI with friends/acquaintances?  & Yes, No \\
& SearchAI & Have you actively searched for information about AI?  & Yes, No \\
\midrule
\multirow{3}{*}{\textbf{Regulation}} 
& AIRegulations & Should there be legal requirements for AI?  & Yes, No \\
& HeardEURegulation & Have you heard of the EU AI regulation? & Yes, No \\
& EUAppropriateRegulation & What is your view of the EU AI regulation?  & Appropriate, Too strict, Not strict enough, Don't know \\
\midrule
\textbf{Politics} 
& VoteIntent & Political orientation based on voting intention & Left, Right, Other \\
\midrule
\multirow{8}{*}{\textbf{Perceptions}} 
& AIReduceShortageWorkers & AI will reduce the impact of labor shortages  & Strongly disagree to Strongly agree \\
& AIEasierLife & AI will make everyday life easier  & Strongly disagree to Strongly agree \\
& AIHealtcareBenefit & AI will bring major benefits in healthcare  & Strongly disagree to Strongly agree, Don't know \\
& AIFieldBenefit & I can imagine which job areas benefit from AI  & Strongly disagree to Strongly agree \\
& AIvsHuman & AI will produce content indistinguishable from human work  & Strongly disagree to Strongly agree \\
& AIFalseInfo & AI will lead to greater spread of misinformation & Strongly disagree to Strongly agree \\
& AIUncontrollable & I fear AI will become uncontrollable  & Strongly disagree to Strongly agree \\
& DevelopAI & Do you see AI developments as a risk or opportunity? & Risk, Opportunity, Both\\
\midrule
\multirow{5}{*}{\textbf{Opportunity Themes}} 
& PosWork & AI as beneficial for the labor market, automation, and job creation & Mentioned, Not mentioned\\
& PosHealth & AI as a tool to advance healthcare, medicine, and care services & Mentioned, Not mentioned \\
& PosLife & AI improving daily life, efficiency, and educational contexts & Mentioned, Not mentioned \\
& PosTech & AI associated with innovation, future technology, and research & Mentioned, Not mentioned \\
& PosGeneral & General positive outlook, openness to AI, or hope for risk mitigation & Mentioned, Not mentioned \\
\midrule
\multirow{5}{*}{\textbf{Risk Themes}} 
& RiskJobs & Fears of job loss, redundancy, or dehumanization in the workplace & Mentioned, Not mentioned \\
& RiskLossOfControl & Concerns about losing control, autonomy, or freedom due to AI & Mentioned, Not mentioned \\
& RiskMisuseRegulation & Ethical concerns, potential misuse, and inadequate regulation & Mentioned, Not mentioned\\
& RiskData & Worries about data privacy, surveillance, or misinformation & Mentioned, Not mentioned \\
& RiskSociety & Uncertainty about AI’s societal impact, especially in education & Mentioned, Not mentioned \\
\bottomrule
\end{tabular}}
\caption{List of variables used in the BN analysis, including structured survey responses and grouped thematic codes derived from open-ended justifications about AI. Variables are grouped by conceptual domain: demographics, exposure to AI, perceptions, political orientation, regulatory opinions, and emergent themes reflecting perceived opportunities or risks.}
\label{tab:merged_bn_vars}
\end{table*}

\subsection{Data}

We use data from the survey \textit{Current Questions on AI (June 2023)}, conducted by FORSA on behalf of the Press and Information Office of the German Federal Government and archived by GESIS\citep{pib2024ai}. The dataset consists of interviews with a probability-based sample of 1,506 individuals aged 14 and above, representative of the German-speaking population. Data were collected over three days (June 26–28, 2023) using computer-assisted telephone interviewing.

Respondents were selected using a dual-frame sampling strategy combining landline and mobile numbers, in accordance with standard practice in Germany\citep{pib2024ai}. The survey provides non-aggregated, individual-level responses to a wide array of questions concerning AI. Given the scarcity of recent public datasets with this level of granularity, it offers a valuable snapshot of current public perceptions.

The questionnaire covers interest in and exposure to AI (e.g., media usage, information-seeking, interpersonal conversations); general perceptions of AI as a risk or opportunity; attitudes toward AI regulation, including views on the EU AI Act; beliefs about AI's social impacts (e.g., in healthcare or job automation); and sociodemographic and political background variables, such as age, sex, education, income, and voting intention. Respondents who identified AI as primarily a risk or an opportunity were asked to elaborate in open-ended follow-ups. These answers were then coded by the survey team into discrete variables capturing key thematic concerns, and are included in the released dataset.

\subsection{Data Preprocessing and Feature Engineering}

We harmonized and recoded the original survey variables for consistency. Ordinal variables (e.g., interest in AI, perceived information level) were recoded into labeled four-point scales, while binary exposure items (media use, discussions, information search) were standardized as “Yes” or “No.” Rare responses such as “Don't know” were treated as missing (threshold: $<$50 cases). Voting intention was recoded into a three-level political preference variable (Left, Right, Other), grouping non-voters and unclassifiable answers under “Other.” Observations with missing values in the selected variables were excluded. The  set of variables used is listed in Table~\ref{tab:merged_bn_vars}.

Two additional datasets were created based on whether respondents viewed AI as a risk, an opportunity, or both. Those selecting “opportunity” or “both” provided open-ended responses on perceived benefits, coded into 24 binary indicators by the survey team. These were grouped into five themes: work and automation; health and care; everyday life and education; technological innovation; and general positivity. Similarly, responses from those identifying AI as a “risk” or “both” were coded into 18 binary indicators and grouped into five themes: job displacement and dehumanization; loss of autonomy and control; misuse and regulatory failure; data security and misinformation; and social uncertainty. These grouped thematic variables are also summarized in Table~\ref{tab:merged_bn_vars}.

\subsection{Bayesian Network Modeling}

\subsubsection{Basic Principles of Bayesian Networks.}

A Bayesian network (BN) is a probabilistic graphical model that encodes the joint distribution of a set of random variables using a directed acyclic graph (DAG) \citep{koller2009, pearl1988probabilistic}. Each node represents a variable, and directed edges represent conditional dependencies. The absence of an edge implies a conditional independence relationship, which can be determined via the d-separation criterion \citep{pearl1988probabilistic}. Given a set of $n$ discrete variables $X_1, \dots, X_n$, the distribution defined by a BN factorizes as:
\[
P(X_1, X_2, \dots, X_n) = \prod_{i=1}^n P(X_i \mid \text{Parents}(X_i)),
\]
\noindent
where $\text{Parents}(X_i)$ denotes the parent nodes of $X_i$ in the DAG. This decomposition allows for an efficient representation of high-dimensional distributions when many conditional dependencies are absent.

In our application, all variables are treated as discrete, with ordinal variables modeled as categorical within the BN framework. Conditional probability tables are directly estimated for each configuration of parent variables, without requiring any augmented parameterization.

\subsubsection{Learning Bayesian Networks.}

Both the network structure and the conditional probability tables were learned directly from the data using a fully data-driven approach. Structure learning was carried out using a score-based method that combines the Tabu search algorithm \citep{tsamardinos2006max} with the Akaike Information Criterion (AIC), a well-established metric that balances model fit and complexity. Given a graph structure $G$ and dataset $D$, the AIC score is defined as:
\[
\text{Score}_{\text{AIC}}(G \mid D) = \log P(D \mid G, \hat{\theta}) - d,
\]
where $P(D \mid G, \hat{\theta})$ is the likelihood of the data under the maximum likelihood parameters $\hat{\theta}$, and $d$ is the number of free parameters in the model. For discrete variables, the log-likelihood can be written as:
\[
\log P(D \mid G, \hat{\theta}) = \sum_{i=1}^{n} \sum_{j=1}^{q_i} \sum_{k=1}^{r_i} N_{ijk} \log \frac{N_{ijk}}{N_{ij}},
\]
where $X_i$ is a node with $r_i$ possible states and $q_i$ parent configurations, and $N_{ijk}$ is the number of observations where $X_i = x_k$ and its parents are in configuration $x_j$. The AIC is particularly suitable for our setting, offering a more permissive penalty than BIC, which is appropriate given the moderate sample size and number of variables \citep{scutari2010learning}.

To improve robustness, we applied a non-parametric bootstrap procedure \citep{scutari2013}, generating 2000 resampled datasets and learning a directed acyclic graph (DAG) for each. The final consensus network was obtained using the \textit{averaged.network} method in the \textit{bnlearn} package\citep{scutari2010learning}, which optimizes the inclusion threshold for each edge based on expected predictive accuracy.

To guide learning and enhance interpretability, we enforced a tiered structure using blacklists. Demographic and political variables were set as roots and were not allowed to receive incoming edges from downstream concepts such as exposure, perception, or regulation. AI-related perceptions and regulatory attitudes were allowed to depend on exposure but constrained from influencing earlier tiers.

Once the structure was learned, the conditional probability tables were estimated using Bayesian parameter learning with a uniform Dirichlet prior \citep{heckerman1995learning}. This regularization approach prevents zero-probability estimates in sparse configurations by ensuring that all entries in the conditional distributions remain strictly positive. Given observed counts $N_{ijk}$ for variable $X_i$ taking value $x_k$ with parent configuration $x_j$, the posterior mean estimate is:
\[
\hat{P}(X_i = x_k \mid \text{Pa}(X_i) = x_j) = (N_{ijk} + \alpha)/(N_{ij} + r_i \alpha),
\]
\noindent
where $N_{ij} = \sum_k N_{ijk}$, $r_i$ is the number of states of $X_i$, and $\alpha = 1$ corresponds to an non-informative uniform prior. This estimator shrinks all probabilities away from the extremes, improving robustness to low-frequency patterns. The models are publicly available via the \texttt{bnRep} R package \citep{leonelli2025bnrep}.

\subsubsection{Analyzing the Bayesian Network Model.}

Given the learned BN, we analyze its structure using probabilistic inference and sensitivity methods. These tools help identify key relationships, quantify the influence of individual variables, and assess the robustness of conclusions.

We perform evidence propagation by conditioning on specific variable states and examining the resulting changes in marginal distributions, highlighting directional effects and dependencies. Scenario-based inference is used to simulate joint configurations of interest and analyze downstream consequences on target variables.

To assess global variable importance, we compute Sobol variance indices \citep{ballester2022computing}, which decompose the variance of a target variable into additive contributions from other nodes. These indices measure how much of the uncertainty in a target’s distribution is attributable to variation in a given input. For a target variable $Y$, the first-order Sobol index for input $X_i$  is:
\begin{equation}
S_i = \text{Var}_{X_i}(\mathbb{E}(Y\, |\, X_i))/\text{Var}(Y).
\end{equation}

This quantifies the proportion of output variance explained by $X_i$ alone, marginalizing over all other variables. The resulting scores serve a role analogous to coefficients in regression, but capture both direct and indirect (mediated) influences within the BN structure.

Finally, we conduct local sensitivity analysis by perturbing entries of the conditional probability tables and quantifying their effect on selected target distributions\citep{ballester2023yodo,leonelli2023sensitivity}. This reveals which parameters most strongly influence predictions and supports robustness checks of specific pathways. Together, these analyses offer a layered understanding of the BN’s logic and predictive behavior, enabling interpretable, data-driven insights.

\begin{figure*}
\centering
\includegraphics[width=0.8\textwidth]{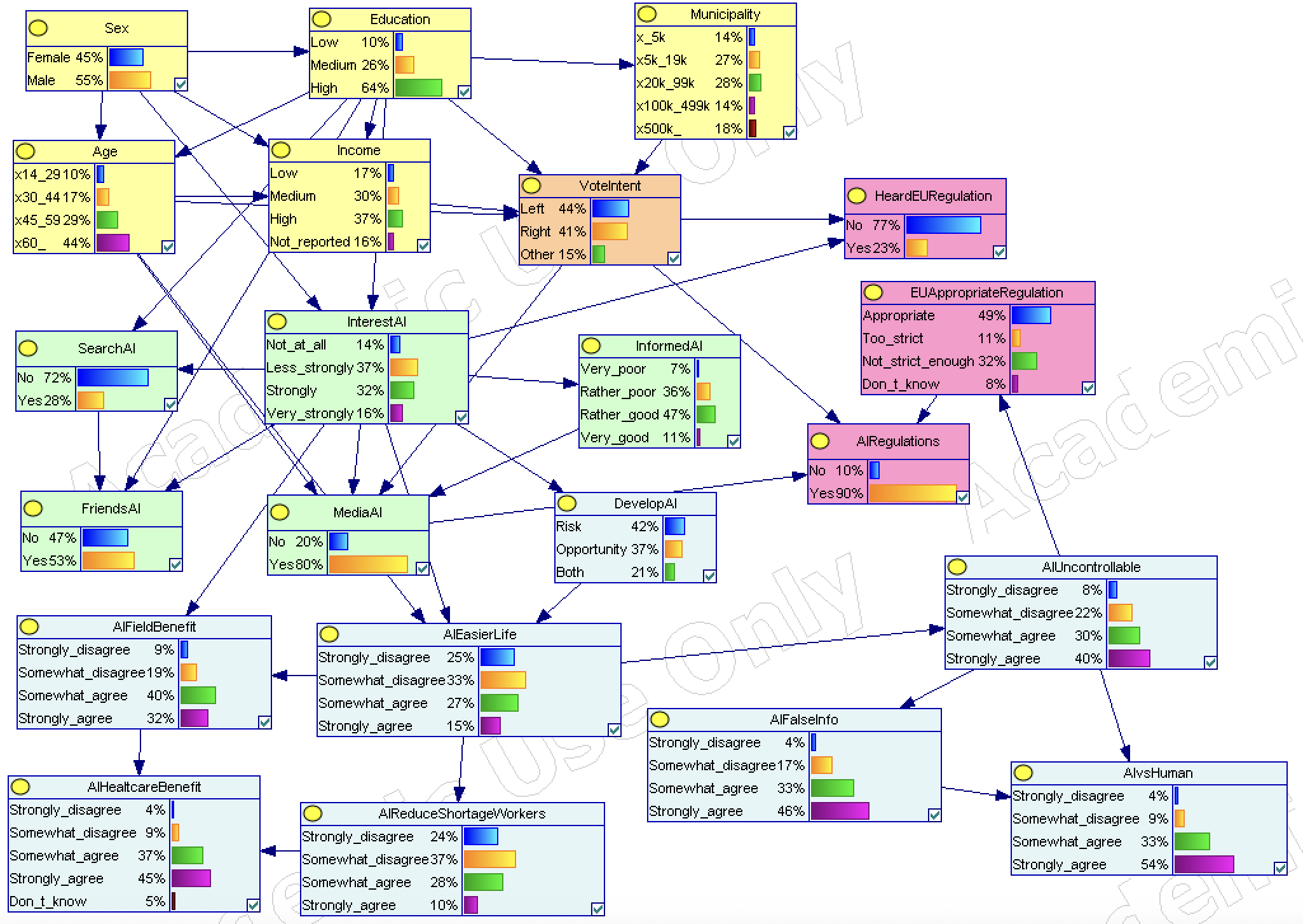}
\caption{Learned BN for the full sample. Nodes are color-coded by variable group: demographics (yellow), exposure (green), regulation (pink), political orientation (orange), and AI perceptions (blue). Visualization produced in GeNIe.}
\label{fig:bn_full}
\end{figure*}

\section{Results}

We begin by presenting the structure and implications of the BN estimated over the full sample of respondents. We then proceed to analyze two submodels learned separately for individuals who primarily perceive AI as a risk and those who see it as an opportunity.

\subsection{Full Sample Bayesian Network}

\subsubsection{Network Structure and Dependencies.}

In the full network (Figure~\ref{fig:bn_full}), \textit{InterestAI} emerges as a central hub, directly influencing both beliefs and information behaviors. It connects to a wide array of downstream nodes, including \textit{DevelopAI} (the core framing of AI as risk/opportunity), as well as \textit{AIEasierLife}, \textit{AIFieldBenefit}, \textit{InformedAI}, \textit{MediaAI}, \textit{FriendsAI}, \textit{SearchAI}, and \textit{HeardEURegulation}. This suggests that personal engagement with AI serves as a primary lens through which both informational exposure and attitudinal formation are structured.

\textit{DevelopAI} itself feeds directly into \textit{AIEasierLife}, which acts as a secondary hub: it shapes beliefs around AI's utility (\textit{AIFieldBenefit}, \textit{AIReduceShortageWorkers}), societal consequences (\textit{AIUncontrollable}), and indirectly, through further downstream effects, beliefs about AI's role in healthcare (\textit{AIHealtcareBenefit}) and misinformation (\textit{AIFalseInfo}, via \textit{AIUncontrollable}). Notably, \textit{AIUncontrollable} links to both \textit{AIFalseInfo} and \textit{AIvsHuman}, and feeds into concerns about regulatory sufficiency through \textit{EUAppropriateRegulation}, which in turn influences \textit{AIRegulations}.

The regulation-related structure features two converging paths. One starts from \textit{InterestAI}, proceeds through \textit{HeardEURegulation}, and terminates in \textit{AIRegulations} via \textit{EUAppropriateRegulation}. A parallel path stems from \textit{VoteIntent}, which exerts direct influence on \textit{MediaAI}, \textit{AIRegulations}, and \textit{HeardEURegulation}. The convergence of these pathways at \textit{AIRegulations} highlights the interplay between perceived institutional exposure and political predispositions in shaping regulatory support.

Demographic variables lie upstream in the structure. \textit{Sex} is a root node influencing \textit{InterestAI}, \textit{Age}, \textit{Education}, and \textit{Income}. \textit{Education} is a pivotal factor, shaping not only \textit{InterestAI}, but also \textit{FriendsAI}, \textit{SearchAI}, \textit{Age}, \textit{Income}, \textit{Municipality}, and \textit{VoteIntent}. \textit{Age} further informs \textit{MediaAI}, \textit{AIEasierLife}, \textit{Income}, and \textit{VoteIntent}. Finally, \textit{Municipality} and \textit{Income} also feed into \textit{VoteIntent}, which acts as a gateway from sociodemographics into the political and attitudinal core of the network.

\subsubsection{Variance Decomposition via Sobol Indices.}

Table~\ref{tab:full_sobol_indices} reports variance-based Sobol indices, which quantify the proportion of output variance attributable to each input variable, for five representative outcomes capturing legislative attitudes (\textit{AIRegulations}, \textit{EUAppropriateRegulation}), AI-related beliefs (\textit{DevelopAI}, \textit{AIUncontrollable}), and policy awareness (\textit{HeardEURegulation}).

Perceptions of AI are primarily shaped by expectations about its usefulness in daily life. The belief that AI will make everyday life easier is the strongest single contributor to how respondents frame AI (49.6\% of variance in \textit{DevelopAI}), followed by self-reported interest in AI (37.3\%). This suggests that optimistic, personally relevant benefits play a dominant role in shaping overall attitudes. Notably, fear that AI may become uncontrollable (\textit{AIUncontrollable}) is itself driven by the same belief about AI's benefits (41.0\%), but also by concerns over misinformation (\textit{AIFalseInfo}, 38.2\%) and the perception that existing regulation is too weak (\textit{EUAppropriateRegulation}, 28.6\%). This combination of optimism and anxiety, mutually reinforcing or counterbalancing, highlights the layered nature of public sentiment toward AI technologies.

Legislative attitudes are influenced through different pathways. Awareness of the EU AI regulation (\textit{HeardEURegulation}) is shaped almost entirely by engagement and information-seeking behaviors, especially interest in AI (70.0\%) and online search activity (15.8\%), with additional influence from political orientation (26.6\%). In contrast, support for legal regulation (\textit{AIRegulations}) depends more on perceived policy adequacy: views on whether the EU AI Act is appropriate explain 32.9\% of variance, followed by political alignment and media exposure. Finally, opinions on whether EU regulation is too strict or too lenient (\textit{EUAppropriateRegulation}) are overwhelmingly driven by techno-anxiety, particularly fear of losing control over AI systems (87.7\%). These results reveal a clear division: perceptual engagement and information-seeking shape awareness, while emotional responses to risk, especially anxiety over control, are central to policy evaluation.

\begin{table*}
\footnotesize
\centering
\renewcommand{\arraystretch}{1.05}
\scalebox{0.6}{
\begin{tabular}{p{4.5cm}ccccc}
\toprule
\textbf{Input Variable} & \textbf{DevelopAI} & \textbf{HeardEURegulation} & \textbf{AIRegulations} & \textbf{EUAppropriateRegulation} & \textbf{AIUncontrollable} \\
\midrule
AIEasierLife & \textbf{49.6}\% & 9.5\% & 0.0\% & \textbf{10.6\%} & \textbf{41.0\%} \\
InterestAI & \textbf{37.3\%} & \textbf{70.0\%} & 0.5\% & 1.3\% & 5.6\% \\
AIFieldBenefit & \textbf{11.3\%} & 8.5\% & 0.0\% & 1.2\% & 4.8\% \\
SearchAI & 7.1\% & \textbf{15.8\%} & 0.0\% & 0.3\% & 1.2\% \\
FriendsAI & 6.2\% & \textbf{11.6\%} & 0.1\% & 0.2\% & 1.0\% \\
AIUncontrollable & 6.1\% & 1.1\% & 0.4\% & \textbf{87.7\%} & -- \\
InformedAI & 5.6\% & \textbf{11.7\%} & 0.5\% & 0.2\% & 0.9\% \\
AIReduceShortageWorkers & 4.5\% & 0.8\% & 0.0\% & 0.9\% & 3.4\% \\
Education & 2.4\% & 6.2\% & 0.3\% & 0.1\% & 0.6\% \\
AIHealtcareBenefit & 1.9\% & 1.1\% & 0.0\% & 0.3\% & 1.1\% \\
HeardEURegulation & 1.8\% & -- & 0.3\% & 0.1\% & 0.3\% \\
MediaAI & 1.8\% & 4.7\% & \textbf{12.3\%} & 0.0\% & 0.1\% \\
Sex & 0.9\% & 1.7\% & 0.0\% & 0.0\% & 0.2\% \\
AIFalseInfo & 0.7\% & 0.1\% & 0.0\% & 8.4\% & \textbf{38.2\%} \\
EUAppropriateRegulation & 0.6\% & 0.1\% & \textbf{32.9\%} & -- & \textbf{28.6\%} \\
AIvsHuman & 0.3\% & 0.1\% & 0.0\% & 5.0\% & \textbf{13.3\%} \\
Income & 0.2\% & 1.8\% & 0.6\% & 0.0\% & 0.1\% \\
Age & 0.2\% & 0.3\% & 0.5\% & 0.6\% & 2.3\% \\
VoteIntent & 0.1\% & \textbf{26.6\%} & \textbf{13.4\%} & 0.0\% & 0.0\% \\
Municipality & 0.0\% & 0.7\% & 0.2\% & 0.0\% & 0.0\% \\
DevelopAI & -- & 3.9\% & 0.0\% & 1.7\% & 7.2\% \\
\bottomrule
\end{tabular}}
\caption{Variance-based Sobol indices (percentage of output variance explained) for each input variable and five target outcomes in the BN model. Bold values indicate variables that explain more than 10\% of the variance for the corresponding output. Dashes (--) indicate the variable was the target and thus excluded from its own analysis.}
\label{tab:full_sobol_indices}
\end{table*}

\subsubsection{Conditional Probabilities and Belief Dynamics.}

We next examine how input variables with an explanatory contribution above 10\% influence the distribution of key target variables when entered as evidence. Tables~\ref{tab:conditional_probs_developai_cleaned}–\ref{tab:conditional_probs_aiuncontrollable_cleaned} display conditional probability tables for the five main outcomes of interest, showing how their distributions shift across relevant predictor categories.

As shown in Table~\ref{tab:conditional_probs_developai_cleaned}, personal attitudes and perceived utility are key drivers of how respondents interpret AI developments. Those who believe that AI will make life easier are substantially more likely to frame it as an opportunity (e.g., 67.9\% for ``Strongly agree'') than as a risk (14.6\%). A similar gradient appears for interest in AI: respondents reporting very strong interest choose ``Opportunity'' (51.0\%) or ``Both'' (23.7\%) far more often than ``Risk'' (25.3\%). These shifts reinforce the earlier variance decomposition, which identified \textit{AIEasierLife} and \textit{InterestAI} as the strongest contributors to variance in \textit{DevelopAI}. At the opposite end, respondents who strongly disagree with the idea that AI will ease daily life are more than 65\% likely to view it as a risk, confirming that techno-optimism plays a protective role against risk-based framing.

\begin{table}
\footnotesize
\centering
\renewcommand{\arraystretch}{1.05}
\scalebox{0.7}{
\begin{tabular}{llccc}
\toprule
Evidence Variable &    Evidence Value &   Both & Opportunity &   Risk \\
\midrule
 Baseline &  & 0.2143 &      0.3666 & 0.4191 \\\midrule
    \multirow{4}{*}{AIEasierLife} & Strongly disagree & 0.1819 & 0.1534 & 0.6648 \\
    & Somewhat disagree & 0.2420 & 0.2713 & 0.4867 \\
    & Somewhat agree & 0.2309 & 0.5133 & 0.2558 \\
    & Strongly agree & 0.1753 & 0.6785 & 0.1463 \\\midrule
  \multirow{4}{*}{AIFieldBenefit} 
    & Strongly disagree & 0.1782 & 0.2314 & 0.5904 \\
    & Somewhat disagree & 0.2035 & 0.2925 & 0.5040 \\
    & Somewhat agree & 0.2223 & 0.3727 & 0.4050 \\
    & Strongly agree & 0.2209 & 0.4421 & 0.3370 \\\midrule
  \multirow{4}{*}{InterestAI} 
    & Not at all & 0.1340 & 0.2095 & 0.6565 \\
    & Less strongly & 0.1823 & 0.3600 & 0.4577 \\
    & Strongly & 0.2761 & 0.3724 & 0.3516 \\
    & Very strongly & 0.2372 & 0.5101 & 0.2527 \\
\bottomrule
\end{tabular}}
\caption{Conditional probabilities for DevelopAI, grouped by evidence. Baseline shown in the first row. }
\label{tab:conditional_probs_developai_cleaned}
\end{table}

\begin{table}
\footnotesize
\centering
\renewcommand{\arraystretch}{1.05}
\scalebox{0.7}{
\begin{tabular}{llcc}
\toprule
              Evidence Variable &    Evidence Value &     No &    Yes \\
\midrule
                       Baseline &                   & 0.7741 & 0.2259 \\
\midrule
     \multirow{2}{*}{FriendsAI} &                No & 0.8167 & 0.1833 \\
                                &               Yes & 0.7360 & 0.2640 \\
\midrule   
    \multirow{4}{*}{InformedAI} &         Very poor & 0.8593 & 0.1407 \\
                                &       Rather poor & 0.7977 & 0.2023 \\
                                &       Rather good & 0.7640 & 0.2360 \\
                                &         Very good & 0.6839 & 0.3161 \\
\midrule
    \multirow{4}{*}{InterestAI} &        Not at all & 0.9317 & 0.0683 \\
                                &     Less strongly & 0.8204 & 0.1796 \\
                                &          Strongly & 0.7382 & 0.2618 \\
                                &     Very strongly & 0.5980 & 0.4020 \\
\midrule
      \multirow{2}{*}{SearchAI} &                No & 0.8031 & 0.1969 \\
                                &               Yes & 0.6981 & 0.3019 \\
\midrule
    \multirow{3}{*}{VoteIntent} &              Left & 0.7091 & 0.2909 \\
                                &             Right & 0.8072 & 0.1928 \\
                                &             Other & 0.8717 & 0.1283 \\
\bottomrule
\end{tabular}}
\caption{Conditional probabilities for HeardEURegulation, grouped by evidence. Baseline shown in the first row. }
\label{tab:conditional_probs_heardeuregulation_cleaned}
\end{table}

\begin{table}
\footnotesize
\centering
\renewcommand{\arraystretch}{1.05}
\scalebox{0.7}{
\begin{tabular}{llcc}
\toprule
                       Evidence Variable &    Evidence Value &     No &    Yes \\
\midrule
                                Baseline &                   & 0.0996 & 0.9004 \\
\midrule
\multirow{4}{*}{EUAppropriateRegulation} & Not strict enough & 0.0658 & 0.9342 \\
                                         &       Appropriate & 0.0776 & 0.9224 \\
                                         &        Too strict & 0.2712 & 0.7288 \\
                                         &        Don't know & 0.1242 & 0.8758 \\
\midrule
                \multirow{2}{*}{MediaAI} &                No & 0.1773 & 0.8227 \\
                                         &               Yes & 0.0803 & 0.9197 \\
\midrule 
             \multirow{3}{*}{VoteIntent} &              Left & 0.0559 & 0.9441 \\
                                         &             Right & 0.1236 & 0.8764 \\
                                         &             Other & 0.1606 & 0.8394 \\
\bottomrule
\end{tabular}}
\caption{Conditional probabilities for AIRegulations, grouped by evidence. Baseline shown in the first row. }
\label{tab:conditional_probs_airegulations_cleaned}
\end{table}

\begin{table*}
\footnotesize
\centering
\renewcommand{\arraystretch}{1.05}
\scalebox{0.7}{
\begin{tabular}{llcccc}
\toprule
                Evidence Variable &    Evidence Value & Not strict enough & Appropriate & Too strict & Don't know \\
\midrule
                         Baseline &                   &            0.3180 &      0.4892 &     0.1143 &     0.0785 \\
\midrule
    \multirow{4}{*}{AIEasierLife} & Strongly disagree &            0.3857 &      0.4307 &     0.1037 &     0.0799 \\
                                  & Somewhat disagree &            0.3340 &      0.4804 &     0.1081 &     0.0775 \\
                                  &    Somewhat agree &            0.2687 &      0.5316 &     0.1222 &     0.0776 \\
                                  &    Strongly agree &            0.2570 &      0.5313 &     0.1318 &     0.0799 \\
\midrule
\multirow{4}{*}{AIUncontrollable} & Strongly disagree &            0.1652 &      0.5456 &     0.1960 &     0.0932 \\
                                  & Somewhat disagree &            0.1397 &      0.6449 &     0.1397 &     0.0756 \\
                                  &    Somewhat agree &            0.2785 &      0.5402 &     0.1087 &     0.0725 \\
                                  &    Strongly agree &            0.4780 &      0.3528 &     0.0878 &     0.0815 \\
\bottomrule
\end{tabular}}
\caption{Conditional probabilities for EUAppropriateRegulation, grouped by evidence. Baseline shown in the first row. }
\label{tab:conditional_probs_euappropriateregulation_cleaned}
\end{table*}

\begin{table*}
\footnotesize
\centering
\renewcommand{\arraystretch}{1.05}
\scalebox{0.68}{
\begin{tabular}{llrrrr}
\toprule
                       Evidence Variable &    Evidence Value & Strongly disagree & Somewhat disagree & Somewhat agree & Strongly agree \\
\midrule
                                Baseline &                   &            0.0804 &            0.2214 &         0.3003 &         0.3979 \\
\midrule
           \multirow{4}{*}{AIEasierLife} & Strongly disagree &            0.0624 &            0.0952 &         0.2033 &         0.6390 \\
                                         & Somewhat disagree &            0.0278 &            0.1890 &         0.3576 &         0.4256 \\
                                         &    Somewhat agree &            0.0945 &            0.3238 &         0.3521 &         0.2296 \\
                                         &    Strongly agree &            0.2056 &            0.3221 &         0.2389 &         0.2334 \\
\midrule
            \multirow{4}{*}{AIFalseInfo} & Strongly disagree &            0.3517 &            0.2927 &         0.1680 &         0.1876 \\
                                         & Somewhat disagree &            0.1447 &            0.3405 &         0.2923 &         0.2226 \\
                                         &    Somewhat agree &            0.0672 &            0.2486 &         0.3896 &         0.2946 \\
                                         &    Strongly agree &            0.0419 &            0.1504 &         0.2498 &         0.5579 \\
\midrule
              \multirow{4}{*}{AIvsHuman} & Strongly disagree &            0.1551 &            0.2010 &         0.2013 &         0.4425 \\
                                         & Somewhat disagree &            0.1344 &            0.3340 &         0.2712 &         0.2604 \\
                                         &    Somewhat agree &            0.0661 &            0.2818 &         0.3898 &         0.2624 \\
                                         &    Strongly agree &            0.0746 &            0.1675 &         0.2584 &         0.4995 \\
\midrule
\multirow{4}{*}{EUAppropriateRegulation} &       Appropriate &            0.0897 &            0.2919 &         0.3316 &         0.2869 \\
                                         &        Too strict &            0.1379 &            0.2708 &         0.2858 &         0.3056 \\
                                         & Not strict enough &            0.0418 &            0.0973 &         0.2630 &         0.5979 \\
                                         &        Don t know &            0.0955 &            0.2135 &         0.2777 &         0.4133 \\
\bottomrule
\end{tabular}}
\caption{Conditional probabilities for AIUncontrollable, grouped by evidence. Baseline shown in the first row. }
\label{tab:conditional_probs_aiuncontrollable_cleaned}
\end{table*}

As shown in Table~\ref{tab:conditional_probs_heardeuregulation_cleaned}, the strongest contrasts in regulatory awareness (\textit{HeardEURegulation}) emerge between individuals with differing levels of AI interest and information engagement. Those with no interest in AI have a 93.2\% probability of being unaware of EU regulations, compared to just 59.8\% among those very interested. Awareness also rises with self-reported knowledge (\textit{InformedAI}) and discussion with peers (\textit{FriendsAI}). 

Table~\ref{tab:conditional_probs_airegulations_cleaned} shows that support for AI-specific legal regulation (\textit{AIRegulations}) is primarily shaped by policy evaluations: respondents who view current regulation as ``Not strict enough'' support legal requirements at a rate of 93.4\%, while those who believe it is ``Too strict'' show only 72.9\% support. This suggests that perceptions of regulatory adequacy serve as a key driver of willingness to endorse legal constraints.

Concerns about AI becoming uncontrollable (\textit{AIUncontrollable}) reflect a layered interplay of optimism, anxiety, and policy evaluation. As shown in Table~\ref{tab:conditional_probs_aiuncontrollable_cleaned}, individuals with strong techno-optimism (\textit{AIEasierLife: Strongly agree}) are less likely to express high concern (23.3\% for ``Strongly agree'') than those who reject AI’s usefulness (63.9\%). However, the most influential factors stem from techno-anxiety: respondents who strongly believe AI spreads misinformation (\textit{AIFalseInfo: Strongly agree}) have a 55.8\% probability of fearing AI is uncontrollable. Similarly, those who perceive current regulation as ``Not strict enough'' are 59.8\% likely to express strong concern, compared to just 28.7\% among those who see it as ``Appropriate''.

\subsubsection{Joint Scenario Reasoning.}

\begin{table*}
\footnotesize
\centering
\renewcommand{\arraystretch}{1.05}
\scalebox{0.7}{
\begin{tabular}{p{5.1cm}p{13.4cm}}
\toprule
\textbf{Scenario Name} & \textbf{Definition (key variables)} \\
\midrule
Baseline (No Evidence) & No evidence provided; full marginal uncertainty retained. \\
Young Informed Left & Age: 14–29; Education: High; Vote: Left; InterestAI: Very strongly; InformedAI: Very good; All exposure variables: Yes. \\
Older Uninformed Right & Age: 60+; Education: Low; Vote: Right; InterestAI: Less strongly; InformedAI: Very poor; All exposure variables: No. \\
Middle Educated Moderate & Age: 45–59; Education: Medium; Vote: Other; InterestAI: Less strongly; InformedAI: Rather poor; MediaAI: Yes; FriendsAI, SearchAI: No. \\
Highly Interested Low Info & Age: 30–44; Education: Medium; Vote: Left; InterestAI: Very strongly; InformedAI: Very poor; All exposure variables: Yes. \\
High Info Low Interest & Age: 30–44; Education: High; Vote: Right; InterestAI: Not at all; InformedAI: Very good; MediaAI: Yes; FriendsAI, SearchAI: No. \\
Urban Conservative & Age: 30–44; Education: Medium; Vote: Right; InterestAI: Strongly; InformedAI: Rather good; All exposure variables: Yes. \\
Rural Progressive & Age: 45–59; Education: High; Vote: Left; InterestAI: Strongly; InformedAI: Rather good; MediaAI, FriendsAI: Yes; SearchAI: No. \\
Young Apathetic & Age: 14–29; Education: Low; Vote: Other; InterestAI: Not at all; InformedAI: Very poor; All exposure variables: No. \\
Engaged Moderate & Age: 30–44; Education: Medium; Vote: Other; InterestAI: Strongly; InformedAI: Rather good; All exposure variables: Yes. \\
Disengaged Elder & Age: 60+; Education: Medium; Vote: Other; InterestAI: Not at all; InformedAI: Rather poor; All exposure variables: No. \\
\bottomrule
\end{tabular}}
\caption{Definition of the ten scenario profiles used in multi-variable evidence analysis.}
\label{tab:scenario_definitions}
\end{table*}

BNs support joint reasoning under multiple sources of evidence. Instead of conditioning on a single variable, we define realistic \emph{scenarios} by fixing combinations of age, education, interest in AI, political orientation, and information exposure. Table~\ref{tab:scenario_definitions} presents ten illustrative profiles, ranging from the \emph{Young Informed Left} to the \emph{Disengaged Elder}, to compare attitudes across diverse population segments.

\begin{figure*}
\centering
\includegraphics[width=0.48\textwidth]{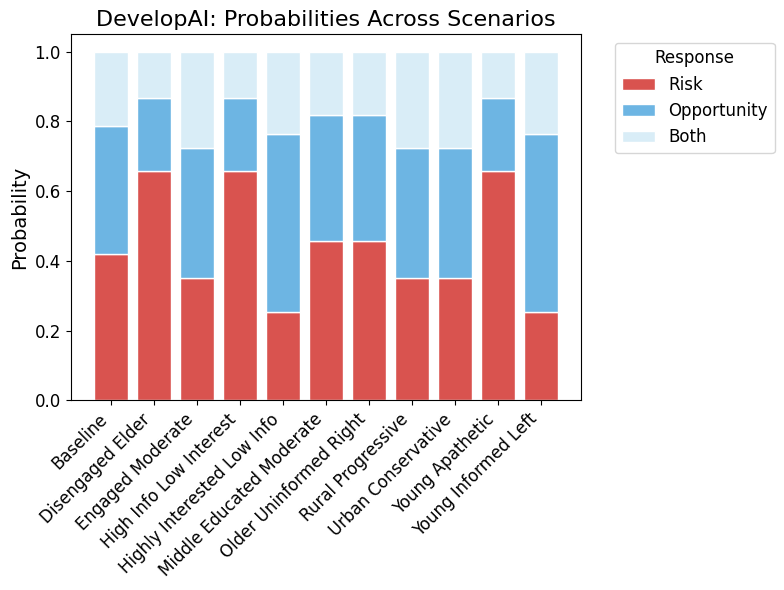}
\includegraphics[width=0.48\textwidth]{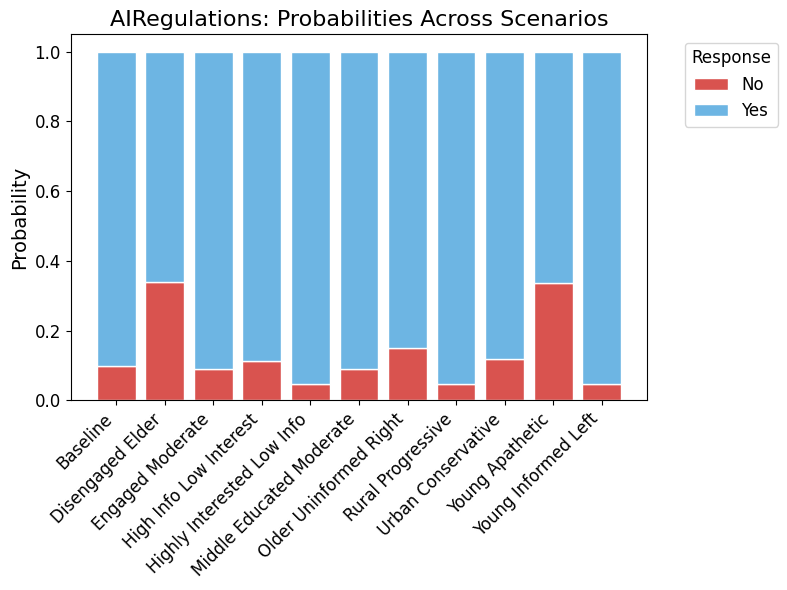}
\includegraphics[width=0.48\textwidth]{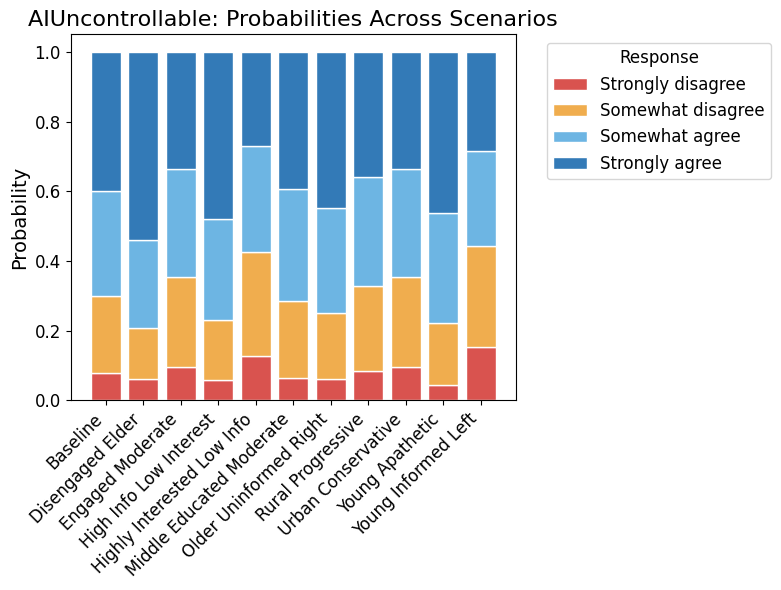}
\caption{Posterior distributions across the ten scenarios for three key target variables: \textit{DevelopAI}, \textit{AIRegulations}, and \textit{AIUncontrollable}.}
\label{fig:scenario_bars}
\end{figure*}

Figure~\ref{fig:scenario_bars} shows the predicted probabilities for three central variables. The \textit{DevelopAI} plot reveals how optimistic beliefs (viewing AI as an opportunity) are prevalent in highly informed and interested groups, while apathy and disengagement drive risk-dominant perceptions. Support for legal regulation of AI (\textit{AIRegulations}) is remarkably high overall but varies modestly with political and engagement variables, with \emph{Young Apathetic} and \emph{Disengaged Elder} profiles showing more skepticism. In contrast, concern over loss of control (\textit{AIUncontrollable}) shows wide variation: techno-anxiety peaks in disengaged and older respondents and subsides among those with balanced or moderate exposure. 

\subsubsection{Parametric Sensitivity Analysis.}

To identify which upstream nodes most strongly influence each target, we conducted a parametric sensitivity analysis alongside the Sobol variance decomposition. For conciseness, sample results are shown in Appendix Figures~\ref{fig:sens_colored_network}–\ref{fig:tornado_heard}, including DAGs colored by local sensitivity values and a tornado plot for \textit{HeardEURegulation}. This analysis quantifies how small changes in the conditional probabilities of ancestral nodes affect downstream distributions. For example, \textit{DevelopAI} is most sensitive to \textit{InterestAI}, with additional effects from \textit{Sex} and \textit{Education}; \textit{HeardEURegulation} is highly influenced by \textit{InterestAI} and \textit{VoteIntent}; and \textit{AIRegulations} is primarily driven by \textit{EUAppropriateRegulation}. 

\subsection{Subpopulation Networks: Risk and Opportunity}

Both subpopulation models are estimated on the same backbone of demographic, exposure, and policy-opinion variables as the full network, but the detailed AI-perception block now consists of the \emph{grouped thematic variables} derived from open-ended answers.

\begin{figure*}
\centering
\includegraphics[width=0.8\textwidth]{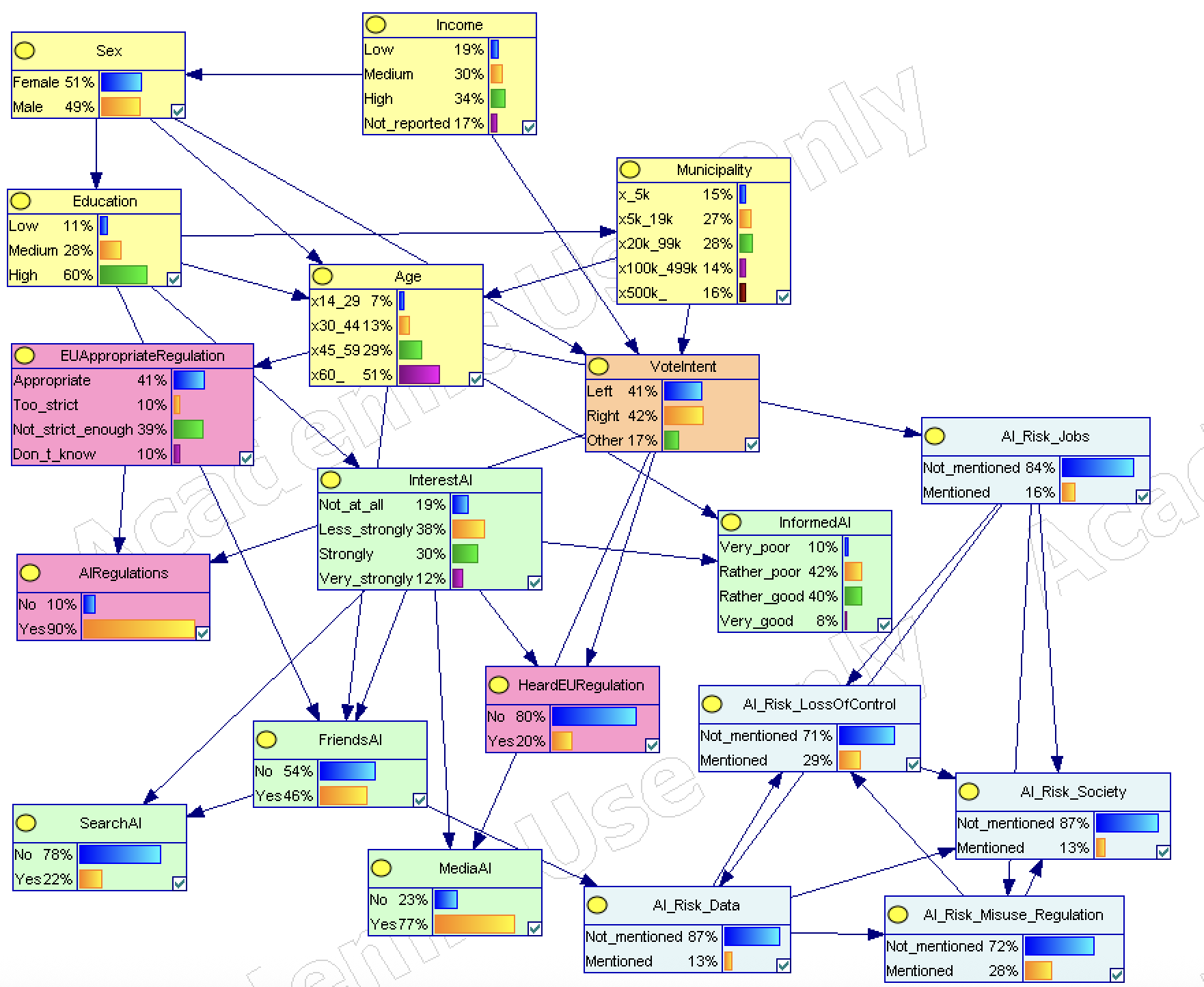}
\caption{BN learned from respondents who view AI as a risk or both a risk and an opportunity, including grouped thematic variables capturing perceived risks. Nodes are color-coded by variable group: demographics (yellow), exposure (green),
regulation (pink), political orientation (orange), and AI perceptions (blue). Visualization produced in GeNIe.}
\label{fig:risk_bn}
\end{figure*}

\subsubsection{Network Structure and Dependencies.}

In the \textit{risk-oriented} network (Figure~\ref{fig:risk_bn}), vote intention (\textit{VoteIntent}) directly influences both regulatory awareness (\textit{HeardEURegulation}) and support for legal requirements (\textit{AIRegulations}), as well as media exposure (\textit{MediaAI}), confirming its central role in shaping both institutional opinions and information behavior. The link from perceived adequacy of the EU regulation (\textit{EUAppropriateRegulation}) to \textit{AIRegulations} remains, while all exposure-related arcs into the regulation block, such as \textit{MediaAI} $\rightarrow$ \textit{AIRegulations} in the full model, disappear. Within the thematic risk layer, all five grouped variables form a tightly connected subgraph, with \textit{AI\_Risk\_Jobs} acting as the main source node influencing the others either directly or via short chains. Demographic and exposure variables have limited direct access to the perception layer, with only \textit{Age} and \textit{FriendsAI} entering this block through \textit{AI\_Risk\_Jobs} and \textit{AI\_Risk\_Data}, respectively.

\begin{figure*}
\centering
\includegraphics[width=0.8\textwidth]{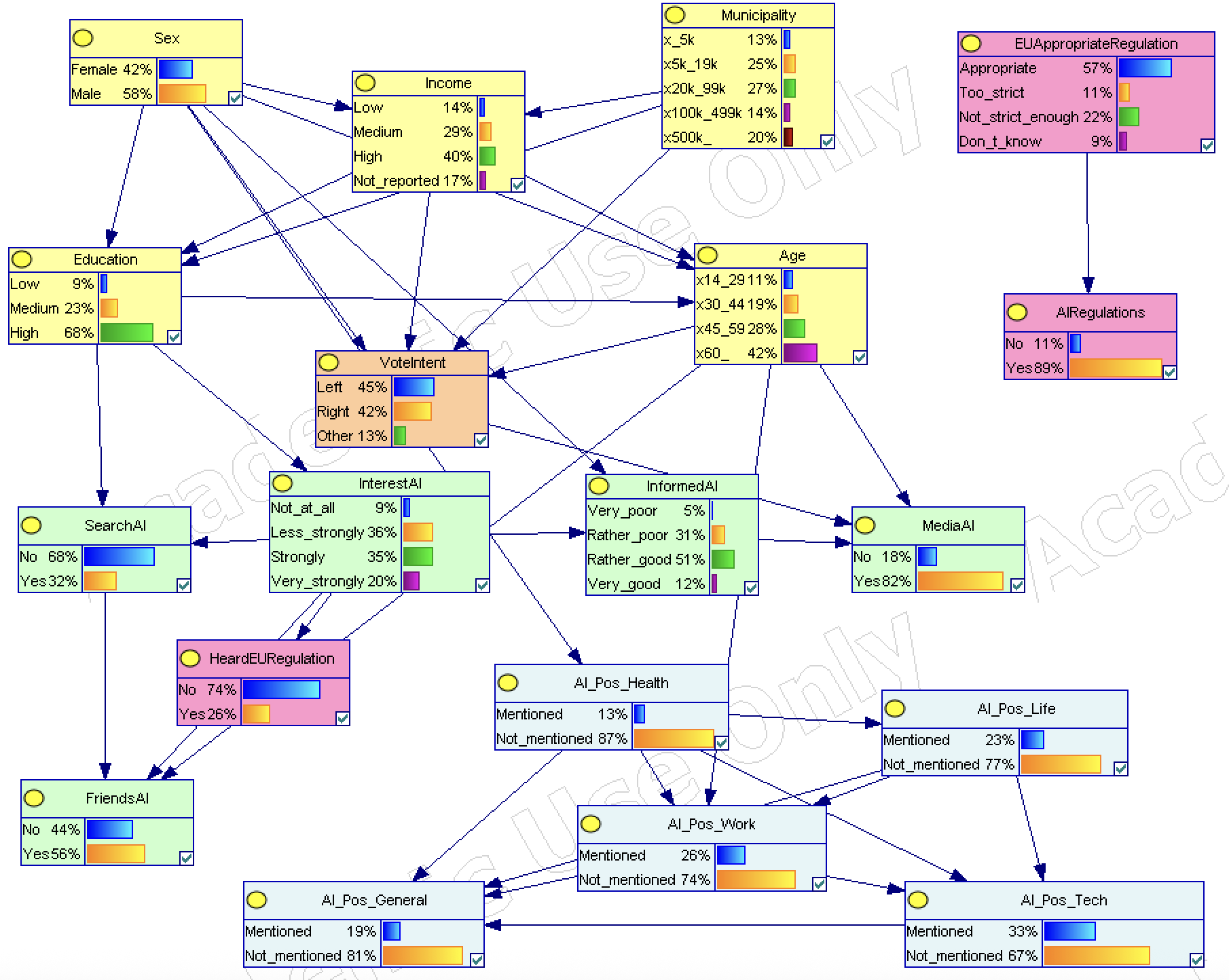}
\caption{BN learned from respondents who view AI as an opportunity or both an opportunity and a risk, including grouped thematic variables capturing perceived benefits. Nodes are color-coded by variable group: demographics (yellow), exposure (green),
regulation (pink), political orientation (orange), and AI perceptions (blue). Visualization produced in GeNIe.}
\label{fig:bn_opportunity}
\end{figure*}

In the \textit{opportunity-oriented} network (Figure~\ref{fig:bn_opportunity}), vote intention still affects \textit{MediaAI}, but no longer connects directly to regulatory support. Instead, support for AI regulation (\textit{AIRegulations}) depends exclusively on \textit{EUAppropriateRegulation}, suggesting that among optimistic respondents, support for legal constraints is shaped more by evaluations of policy content than by political alignment. The grouped benefit themes form a dense, layered structure with \textit{AI\_Pos\_Health} and \textit{AI\_Pos\_Life} acting as upstream nodes. These flow into \textit{AI\_Pos\_Work} and \textit{AI\_Pos\_Tech}, which in turn feed into the most general category (\textit{AI\_Pos\_General}). In contrast to the risk network, the thematic benefit nodes receive several direct inputs from demographic variables: \textit{Sex} enters \textit{AI\_Pos\_Health}, and \textit{Age} influences \textit{AI\_Pos\_Work}. Exposure and engagement variables such as \textit{InterestAI}, \textit{InformedAI}, \textit{FriendsAI}, and \textit{SearchAI} are again structured hierarchically but do not directly connect to the perception nodes. The paths into regulation mirror a technocratic progression: \textit{InterestAI} drives awareness (\textit{HeardEURegulation}) via standard exposure channels, while \textit{EUAppropriateRegulation} alone governs support.

Together, these two networks reveal how overall framing (risk vs. opportunity) restructures the dependencies among background factors, exposure, and regulation, as well as the internal logic of belief formation. While the risk model is politically activated and thematically entangled, the opportunity one is demographically modulated and structurally layered.

\subsubsection{Variance Decomposition via Sobol Indices.}

Table \ref{tab:sobol_risk_opportunity_final} reports the first-order Sobol indices within the risk and opportunity subnetworks, focusing again on the three regulation-related outcomes and one output from each group of specific constructs: \textit{AI\_Risk\_Jobs} for the risk subnetwork and \textit{AI\_Pos\_Health} for the opportunity subnetwork. Outputs that were disconnected from their corresponding subnetwork, such as \textit{AIRegulations} and \textit{EUAppropriateRegulation} in the opportunity case, were not included since the Sobol indices are exactly zero for all inputs.

For \textit{HeardEURegulation}, we observe a clear contrast between the two subnetworks. In the risk subnetwork, the dominant contributors are \textit{InterestAI} (65.7\%), \textit{SearchAI} (14.6\%), and \textit{VoteIntent} (26.0\%), closely aligned with the full-network result where \textit{InterestAI} and \textit{SearchAI} also rank among the most influential. In the opportunity subnetwork, the variance is explained by exposure-related variables: \textit{InterestAI}, \textit{SearchAI}, \textit{FriendsAI}, and \textit{InformedAI}, with no meaningful contribution from demographics or \textit{MediaAI}.

For \textit{AIRegulations}, the most important factor in the risk subnetwork is \textit{EUAppropriateRegulation} (57.1\%), followed by \textit{VoteIntent} (21.4\%). This replicates the overall network structure, where \textit{EUAppropriateRegulation} explains 32.9\% of variance in \textit{AIRegulations}, and \textit{VoteIntent} accounts for 13.4\%. The larger contribution observed here may result from isolating the pathway without interference from other, unrelated nodes. Notably, variables such as \textit{InformedAI} and \textit{FriendsAI} play a minimal role, suggesting that opinions on AI regulation are more strongly tied to institutional and political cues than to general AI familiarity or sentiment.

For \textit{EUAppropriateRegulation}, \textit{Age} dominates the variance in the risk subnetwork (85.1\%), a finding not visible in the full model where \textit{AIEasierLife} and \textit{AIUncontrollable} drive most of the variation. This difference highlights how the removal of highly influential nodes outside the regulation construct can expose underlying demographic patterns that were previously masked. The influence of \textit{InformedAI} and  \textit{AIRegulations} remains modest, consistent with their lower Sobol scores in the full model.

Turning to the specific outputs, no variable explains a large share of the variance. The highest contribution is from \textit{AI\_Risk\_Misuse\_Regulation} (10.6\%) to \textit{AI\_Risk\_Jobs}, indicating that no single input dominates the variance in these thematic outcomes.

\begin{table*}
\footnotesize
\centering
\renewcommand{\arraystretch}{1.05}
\scalebox{0.5}{
\begin{tabular}{lcccc|cc}
\toprule
\textbf{Input Variable} & 
\multicolumn{4}{c|}{\textbf{Risk Subnetwork}} & 
\multicolumn{2}{c}{\textbf{Opportunity Subnetwork}} \\
\cmidrule(lr){2-5} \cmidrule(lr){6-7}
& \textbf{EUAppropriateRegulation} & 
\textbf{AIRegulations} & 
\textbf{HeardEURegulation} & 
\textbf{AI\_Risk\_Jobs} & 
\textbf{HeardEURegulation} & 
\textbf{AI\_Pos\_Health} \\
\midrule
\textbf{Age} & \textbf{85.1\%} & 0.2\% & 0.2\% & 7.6\% & 0.2\% & 0.1\% \\
\textbf{Education} & 2.6\% & 0.0\% & 4.9\% & 0.2\% & 4.8\% & 0.0\% \\
\textbf{Income} & 0.0\% & 0.4\% & 0.7\% & 0.0\% & 0.3\% & 0.1\% \\
\textbf{Municipality} & 1.1\% & 0.6\% & 1.2\% & 0.1\% & 0.1\% & 0.0\% \\
\textbf{Sex} & 0.5\% & 0.0\% & 0.0\% & 0.1\% & 0.0\% & 7.0\% \\
\textbf{VoteIntent} & 0.0\% & \textbf{21.4\%} & \textbf{26.0\%} & 0.0\% & 0.0\% & 0.1\% \\
\textbf{FriendsAI} & 2.3\% & 0.0\% & 9.5\% & 0.2\% & \textbf{13.4\%} & 0.0\% \\
\textbf{InformedAI} & 1.7\% & 0.0\% & 8.9\% & 0.2\% & \textbf{16.9\%} & 0.2\% \\
\textbf{InterestAI} & 0.2\% & 0.0\% & \textbf{65.7\%} & 0.0\% & \textbf{100.0\%} & 0.0\% \\
\textbf{MediaAI} & 0.0\% & 0.4\% & 5.4\% & 0.0\% & 0.4\% & 0.0\% \\
\textbf{SearchAI} & 0.2\% & 0.0\% & \textbf{14.6\%} & 0.0\% & \textbf{21.7\%} & 0.0\% \\
\textbf{AIRegulations} & 1.0\% & -- & 0.5\% & 0.0\% & 0.0\% & 0.0\% \\
\textbf{EUAppropriateRegulation} & -- & \textbf{57.1\%} & 0.0\% & 0.2\% & 0.0\% & 0.0\% \\
\textbf{HeardEURegulation} & 0.0\% & 0.5\% & -- & 0.0\% & -- & 0.0\% \\
\textbf{AI\_Risk\_Jobs} & 1.9\% & 0.0\% & 0.0\% & -- & NA & NA \\
\textbf{AI\_Risk\_Misuse\_Regulation} & 0.1\% & 0.0\% & 0.0\% & \textbf{10.6\%} & NA & NA \\
\textbf{AI\_Risk\_LossOfControl} & 0.1\% & 0.0\% & 0.0\% & 9.4\% & NA & NA \\
\textbf{AI\_Risk\_Society} & 0.1\% & 0.0\% & 0.0\% & 6.0\% & NA & NA \\
\textbf{AI\_Risk\_Data} & 0.0\% & 0.0\% & 0.1\% & 3.2\% & NA & NA \\
\textbf{AI\_Pos\_Work} & NA & NA & NA & NA & 0.0\% & 2.8\% \\
\textbf{AI\_Pos\_Tech} & NA & NA & NA & NA & 0.0\% & 7.0\% \\
\textbf{AI\_Pos\_General} & NA & NA & NA & NA & 0.0\% & 5.5\% \\
\textbf{AI\_Pos\_Life} & NA & NA & NA & NA & 0.0\% & 1.9\% \\
\textbf{AI\_Pos\_Health} & NA & NA & NA & NA & 0.0\% & 1.9\% \\
\bottomrule
\end{tabular}
}
\caption{Variance-based Sobol indices (percentage of output variance explained) for each input variable across target variables in the \textbf{Risk} and \textbf{Opportunity} subnetworks. Bold values indicate variables that explain more than 10\% of output variance. Double dashes (--) indicate that the variable was the target output in that column. \textit{NA} indicates that the variable was not included in the corresponding BN submodel.}
\label{tab:sobol_risk_opportunity_final}
\end{table*}

\subsubsection{Conditional Probabilities and Belief Dynamics.}

We replicate the conditional probability analysis for the three regulation-related outputs in the subnetworks, focusing exclusively on demographic predictors. 

For \textit{HeardEURegulation}, the results from the subnetwork confirm a significant political gradient: respondents who identify as left-leaning are more likely to have heard about EU regulation (26.8\%) compared to the baseline (19.6\%), while right-wing respondents fall slightly below the baseline (15.7\%), and those selecting ``Other'' display the lowest awareness (11.6\%). These patterns align with those reported in Table~\ref{tab:conditional_probs_heardeuregulation_cleaned}, although the magnitude of the contrast is slightly more pronounced in the subnetwork. Notably, this confirms the stable role of political orientation in shaping regulatory awareness, even when isolating regulation-specific constructs.

For \textit{AIRegulations}, we again observe that left-leaning respondents express stronger support for AI regulation (95.4\% compared to a baseline of 90.5\%), whereas right-leaning and ``Other'' respondents express slightly reduced support (87.2\% and 86.7\%, respectively). The direction and magnitude of these effects are highly consistent with the full-network results, reinforcing the interpretation that political stance is a reliable predictor of regulatory preferences.

In the case of \textit{EUAppropriateRegulation}, age stratification reveals substantial heterogeneity. Respondents aged 14--29 are significantly more likely to consider current regulation ``Appropriate'' (70.7\% versus a baseline of 40.9\%), with a corresponding drop in the ``Not strict enough'' category. Middle-aged groups (30--59) align more closely with the baseline, while older respondents (60+) show a slightly lower probability of selecting ``Appropriate'' and a marginal increase in ``Don't know''. These age-related shifts were largely absent from the full-network conditional distributions, where attitudes were primarily shaped by optimism and concern variables. 

\section{Discussion}
\subsection{Patterns in Public Attitudes toward AI}

Our findings reveal that how individuals frame artificial intelligence (as primarily a risk, an opportunity, or both) is strongly shaped by perceived usefulness and personal engagement. The belief that AI will make life easier emerged as the most influential driver of opportunity framing, reinforcing previous research that links techno-optimism to age and education \citep{bergdahl2023, dsit2024wave4}. At the same time, fear that AI may become uncontrollable is associated with heightened concern about misinformation and insufficient regulation, echoing work on techno-anxiety and institutional mistrust \citep{li2020chb, dong2024}. Importantly, these attitudes are not mutually exclusive: many respondents endorse both hopeful and fearful views, supporting recent claims that AI perceptions are multi-dimensional rather than polarized \citep{montag2023, schiavo2024}.

In contrast to models that assume direct ideological effects, our analysis shows that support for legal regulation is largely mediated by respondents’ evaluations of policy adequacy. The variable \texttt{EUAppropriateRegulation} consistently ranked among the most influential predictors of support for legal requirements, suggesting that citizens do not simply support or oppose regulation based on partisan cues, but critically assess whether current efforts go far enough. This finding refines the picture presented by earlier work on regulatory preferences \citep{chiarini2024, laux2023}, indicating that perceived institutional performance, rather than exposure alone, drives willingness to endorse governance interventions.

Engagement with information sources, particularly self-reported interest in AI and active search behavior, played a central role in shaping awareness of EU regulation. Respondents who reported high interest were significantly more likely to have heard of regulatory initiatives, while levels of informedness and interpersonal discussion also contributed. However, these factors did not directly translate into policy support. This confirms findings from prior studies \citep{montag2023, schiavo2024} that awareness is driven by cognitive engagement and information exposure, while support for regulation rests on deeper evaluative and emotional layers, such as trust, perceived risk, and institutional credibility.

The comparison between subpopulation networks, those viewing AI as a risk vs. an opportunity, offers further insight into belief structures. Among risk-oriented respondents, regulatory attitudes are more politically activated, and thematic concerns are tightly interconnected. In this group, political orientation plays a prominent role, consistent with findings from polarized and lower-trust contexts such as Southern Europe and Latin America \citep{gur2024, yarovenko2024}. In contrast, opportunity-oriented respondents display a more demographically driven pattern, where support for regulation flows through assessments of policy content rather than ideology. This suggests that logics of belief formation differ across framing groups: risk attitudes appear more ideologically and emotionally embedded, while opportunity attitudes align more closely with technocratic reasoning.

\subsection{Engaging the Public in Governance}

These findings have important implications for the design of AI governance and public engagement strategies. Scenario-based inference revealed strong attitudinal contrasts between population segments, such as the \emph{Young Informed Left} and the \emph{Disengaged Elder}. Such contrasts suggest that a one-size-fits-all communication strategy is unlikely to succeed. Instead, engagement efforts should be tailored: emphasizing practical benefits and accessibility for disengaged groups, while focusing on procedural adequacy and institutional transparency for politically skeptical individuals. The observed alignment between interest, exposure, and awareness also implies that public education campaigns may be most effective when targeting interest-driven audiences who already engage with technology-related content.

The analysis further suggests that concerns about uncontrollability and policy insufficiency are central to regulatory dissatisfaction, aligning with studies that link fear of AI to weakened trust in governance \citep{li2020chb, chiarini2024}. This highlights the need for risk communication strategies that go beyond factual reassurances and address perceived control, fairness, and legitimacy. Policymakers should recognize that public unease often reflects not just fear of the technology itself, but a broader anxiety about who governs its use and how.

In light of the EU AI Act’s phased implementation, our results underscore the conditional nature of public support. Even individuals who express general agreement with regulation adjust their attitudes based on how adequate current efforts appear. This supports critiques of the Act’s technocratic framing \citep{kop2021, laux2023}, which may fail to resonate with public expectations around fairness, autonomy, and accountability. Making regulatory content accessible and interpretable to the public may be just as important as enforcing compliance or technical standards.

Finally, the modeling framework introduced here provides a practical tool for policymakers seeking to monitor and respond to public opinion. By identifying which population segments are most responsive to particular aspects of regulation (such as awareness campaigns, policy framing, or risk communication) our BN approach enables a more targeted and adaptive form of governance. Future extensions could integrate this model into participatory policymaking platforms or media monitoring systems to support real-time adjustment of engagement strategies.

\subsection{Contributions, Limitations, and Paths Forward}

Methodologically, this study demonstrates the value of probabilistic graphical models for analyzing public attitudes toward emerging technologies. BNs offer a transparent and flexible approach for modeling belief systems, capturing not only direct associations but also conditional dependencies and mediated effects. Unlike traditional regression or structural equation models, BNs accommodate complex interrelations among variables without assuming linearity or unidirectional causality. The use of bootstrapped structure learning, Sobol variance decomposition, and scenario-based inference provides both robustness and interpretability, aligning with recent calls for more transparent and exploratory modeling in high-stakes policy domains \citep{rudin2019}.

Nonetheless, the study has several limitations. First, the analysis is based on cross-sectional data, which limits our ability to draw causal conclusions. While the tiered structure and blacklists reflect plausible causal ordering, they cannot establish temporal precedence. Second, although the dataset is rich, it pertains to a single national context and results may not generalize to other countries without adjustment for institutional and cultural factors. Third, as the network grows in size and complexity, interpretability may diminish. Future studies could explore the use of regularization techniques, sparse priors, or expert elicitation to constrain structure learning and improve scalability.

Several avenues for future research emerge. Longitudinal or repeated cross-sectional data could be used to study belief dynamics over time, particularly in response to policy changes or media events. Comparative studies across different EU countries, or between European and non-European contexts, could reveal how institutional design and public trust interact to shape attitudes. Finally, integrating BNs with qualitative data, such as open-ended survey responses or focus group transcripts, could deepen understanding of the cultural narratives and personal experiences that underpin AI-related beliefs. Such extensions would help further bridge the gap between interpretability, policy relevance, and methodological rigor in the study of technology and society.

\section{Conclusions}

This study presents the first application of a BN to analyze public attitudes toward AI regulation using nationally representative European survey data. By modeling political, psychological, and informational factors within a unified BN framework, we uncover how beliefs co-vary and condition one another, revealing distinct pathways that shape awareness, trust, and support for legal oversight. Our findings emphasize that regulatory attitudes are shaped not only by ideology or exposure but by perceived adequacy and emotional responses to risk. The model’s interpretability enables targeted insights for policymakers, offering a transparent tool for aligning AI governance with public sentiment. Future work can extend this approach across countries and over time to monitor evolving attitudes and inform adaptive regulation in an increasingly AI-integrated society.

\bibliographystyle{elsarticle-harv} 
 \bibliography{bib}






\appendix
\section{Sensitivity Analysis Plots}

\begin{figure}[ht]
\centering
\includegraphics[width=0.8\linewidth]{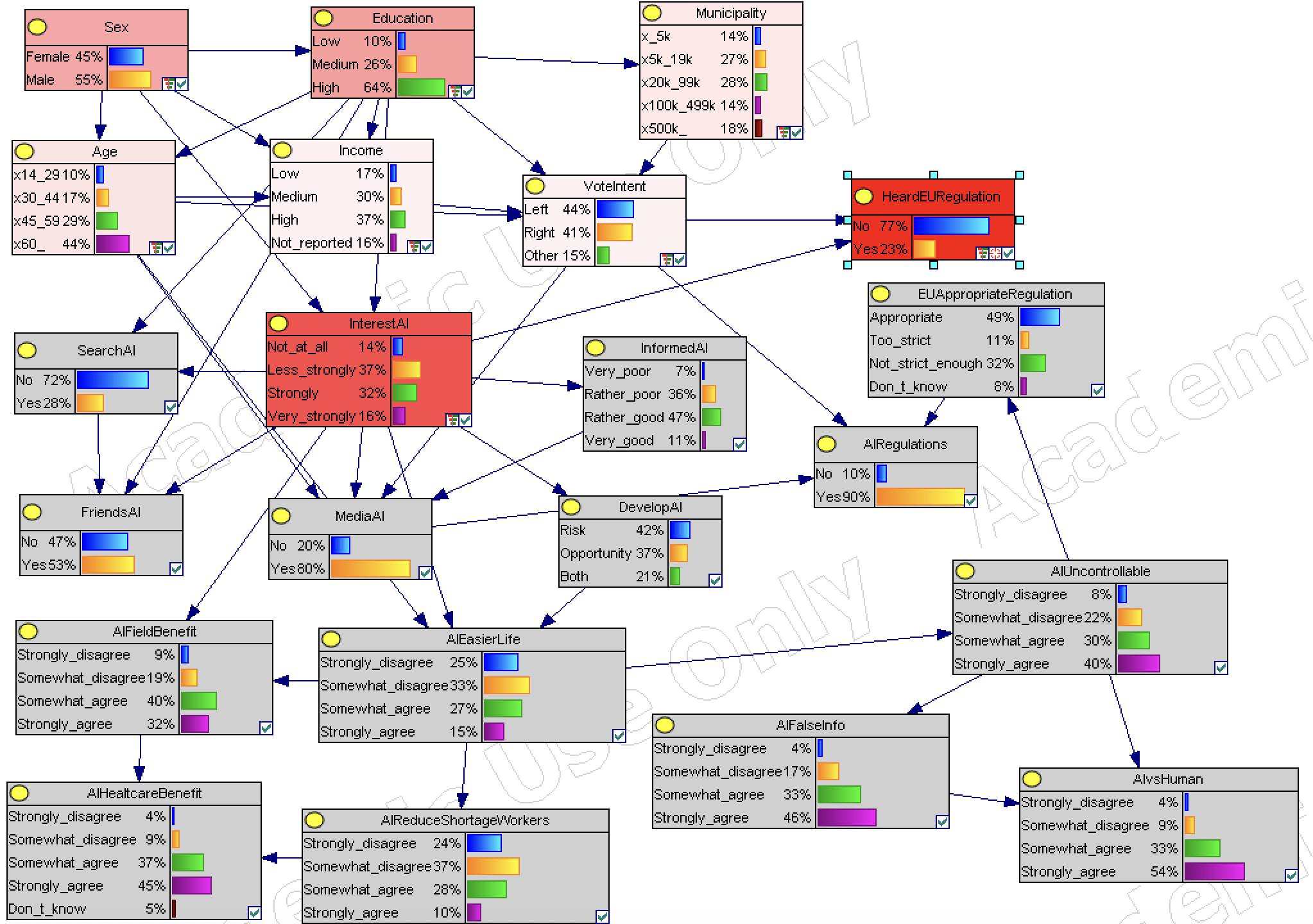}
\caption{Full Bayesian network with nodes shaded by their parametric sensitivity to the variable HeardEURegulation.
Gray nodes have no measurable effect, while nodes shaded in red indicate increasing levels of influence, with darker shades corresponding to stronger effects.}
\label{fig:sens_colored_network}
\end{figure}

\begin{figure}[ht]
\centering
\includegraphics[width=0.7\linewidth]{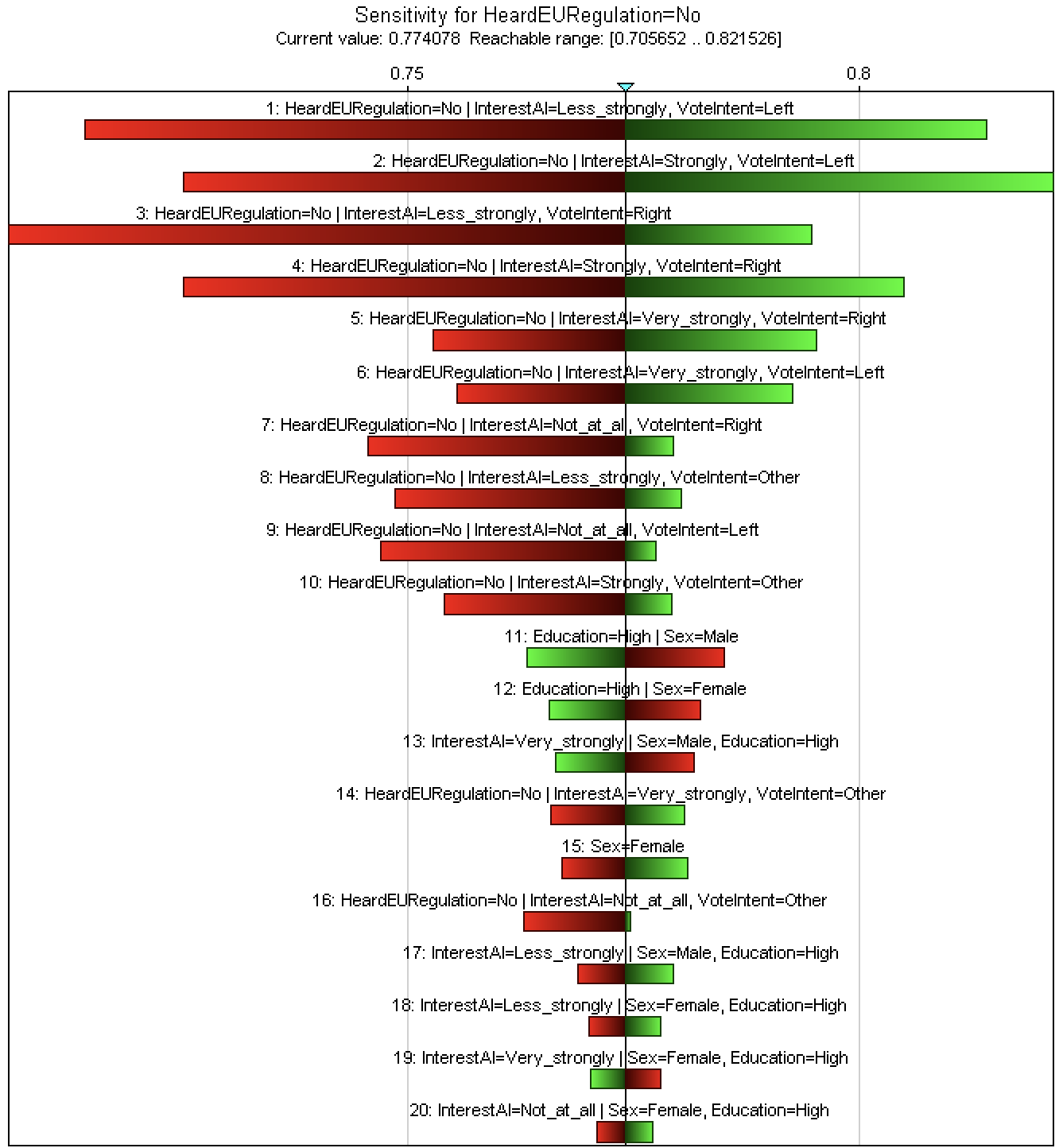}
\caption{Tornado diagram for the variable \textit{HeardEURegulation = No}, showing the effect of changing parameters of upstream nodes. Each bar represents a conditional configuration (e.g., joint levels of \textit{InterestAI} and \textit{VoteIntent}), and its length corresponds to the shift in probability under that scenario. Colors indicate the direction of parameter change: green for increases in the input variable, and red for decreases.}
\label{fig:tornado_heard}
\end{figure}

\end{document}